# Common Reusable Verification Environment for BCA and RTL Models

Giuseppe Falconeri, Walid Naifer, Nizar Romdhane

STMicroelectronics - OCCS (On Chip Communication Systems)

## Abstract

*This paper deals with a common verification methodology and environment for SystemC BCA and RTL models. The aim is to save effort by avoiding the same work done twice by different people and to reuse the same environment for the two design views. Applying this methodology the verification task starts as soon as the functional specification is signed off and it runs in parallel to the models and design development. The verification environment is modeled with the aid of dedicated verification languages and it is applied to both the models. The test suite is exactly the same and thus it's possible to verify the alignment between the two models. In fact the final step is to check the cycle-by-cycle match of the interface behavior. A regression tool and a bus analyzer have been developed to help the verification and the alignment process. The former is used to automate the testbench generation and to run the two test suites. The latter is used to verify the alignment between the two models comparing the waveforms obtained in each run. The quality metrics used to validate the flow are full functional coverage and full alignment at each IP port.*

## 1. Introduction

Nowadays, System On Chips are increasing in terms of complexity and time to market is becoming more and more critical and functional verification is a bottleneck.

Development time of such activity takes 80% of design effort. This effort is mainly spent to test RTL design, since it will be the circuit to map on silicon. The introduction of BCA development in the flow is becoming wider. The fast simulation of BCA models permits to fast find the optimized configuration, in terms of bandwidth, area and power consumption. Therefore, these models are becoming key elements in SoC development and the constraints in terms of functional point view are similar to RTL, that's why it is necessary to have the powerful verification environment also for the models.

Moreover having a common verification environment that is reusable for both BCA and RTL can give a big benefit. The idea of the common verification environment is not new [1] and we want to follow this new strategy because of the gains in terms of development time and accuracy of the verification since the random traffic and automatic checkers can be applied for both the views of the design.

In this paper the common verification environment developed for the dynamic functional verification of the STBus[1] [2] components is described.

## 2. The Past flow

In the past there was no strategy for a common verification of the two views of the IPs. The BCA model verification and the RTL verification were two different activities managed by two or more different teams.

In fact in the verification of the BCA models the test bench was developed by the model owner and occupied a short time of his whole activity. It was based on a very basic model of harnesses written in SystemC and doing write then read operations towards a memory model. The tests cases were directive and allowed checking particular features of the design. And a lot of checks were done visually.

On the opposite the verification of RTL model was based on Verisity Specman [3] tool allowing random generation and automatic checks that cover all functional rules. Whole activity was performed independently from design.

The fact that verification environment was handled by the BCA model owner makes self-error detection more difficult. The test bench was also not strong enough to reach corner cases. Other drawbacks were that the effort spent to develop verification environment was duplicated in RTL and BCA development and there was no way to understand "quality metrics" like coverage for BCA so a new strategy for the verification becomes fundamental.

## 3. STBus Overview

The aim of this section is to introduce the STBus, the communication system developed for System-On-Chip in ST. The STBus is a set of protocols, interfaces and architectural specifications defined to implement the communication network of digital systems such as microcontrollers for different applications (set-top box, digital camera, MPEG decoder, GPS). The STBus protocol consists of three different types (namely Type I, Type II and Type III), each one associated with a different interface and capable of differing performance levels.

---

[1] STMicroelectronics proprietary on-chip bus`

- *Type I* is a simple synchronous handshake protocol with a limited set of available command types, suitable for register access and slow peripherals.
- *Type II* is more efficient than type I because it supports split transactions and pipelining. The transaction set includes simple read/write operation with different sizes (up to 64 bytes) and also specific operations. Transactions may also be grouped together into *chunks* to ensure allocation of the slave and so ensure no interruption of the data stream. It is typically suited for External Memory controllers. A limitation of this protocol is that the traffic must be ordered.
- *Type III* is the most efficient, because it adds support for out-of-order transactions and asymmetric communication (length of request packet different from the length of response packet) on top of what is already provided by Type II. CPUs, multichannel DMAs and DDR controllers can therefore use it.

All the initiators and the targets must have one of these three interfaces and they can communicate each other independently of the type used because type converters into the interconnect (Figure 1) can be used.

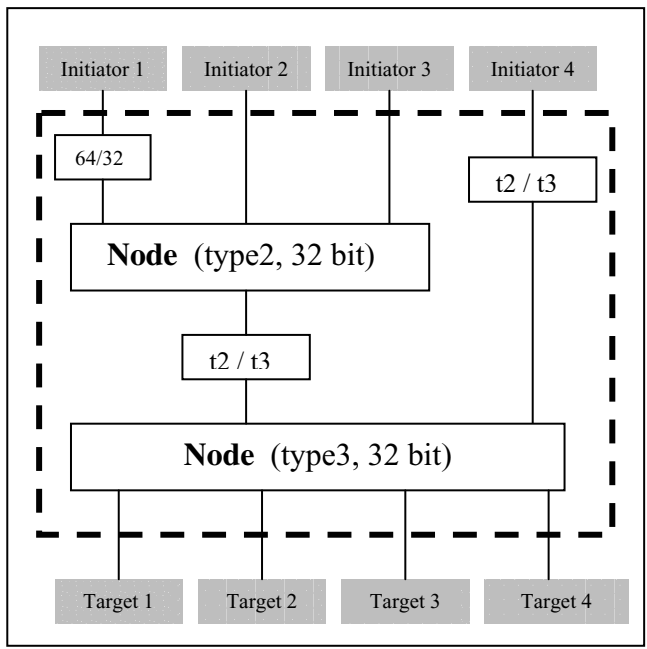

**Figure 1: Example of communication network (interconnect) within a SOC**

The STBUS provides also the size conversion when the initiators and targets have different data bus size. So the STBus is the block to which all initiators and targets must be connected and it performs conversions, arbitration and routing. The arbitration is the process consisting of deciding which initiator, among the ones asking to start a transmission, can take possession of the bus; the latter consists in the propagation of the signals across the bus from an initiator interface to a target interface.

A wide variety of arbitration policies is also available, to help system integrators meet initiators and system requirements. These include bandwidth limitation, latency arbitration, LRU, priority-based arbitration and others.

This is one of the main characteristics of the STBUS. It's not only a single bus or a set of buses, but it can be a hierarchical communication network composed of more than one router. Moreover the various parts of the interconnect can have different width of data bus, different speed and different communication protocol. This can be done connecting a set of 4 basic components: nodes, size converters, type converters and register decoders.

In the following picture an example of an interconnect containing nodes (responsible for routing and arbitration) and some converters is shown.

The architecture of the STBUS is not fixed, but can be different device by device. According to the system requirements it is possible to choose a single shared bus, that gives the better results in terms of wiring congestion and area occupations, but can lead to worse results in terms of performance, or a crossbar (full or partial), that leads better results in terms of performance of the system, but worse results in terms of area and wiring congestion. In this case two different transactions in the same time are possible

## 4. Common Verification Flow

Since the BCA and the RTL models have the same requirements in terms of functional verification (with respect to the specification), it is convenient to have common verification flow for both RTL and BCA. These requirements are:
o Random traffic generation
o Automatic Check on protocol interfaces
o Automatic Check on data integrity: the DUT (Design Under Test) outputs' data correspond to the inputs' one, with respect to the specifications
o Functional and Code Coverage Metrics

In this section the common verification flow used for both BCA and RTL is described. The goal is to have a unique verification environment so that the effort spent to develop it is done only once and not duplicated as in the past. The verification environment development should not depend on the model data type (BCA or RTL). To be efficient, verification activity must be done by a third part, to be independent from other activities. The functional specifications must be the only reference of verification implementation. The BCA or RTL verification activities could also serve to correct verification implementation. In fact some bugs could be given by verification environment.

The common part of the test bench is completely developed in 'e' language and the BCA or the RTL model can be plugged in it. The architecture of the test bench (Figure 2) is standard and it's described in the picture below. All the gray components are written in 'e' code and the DUT can be RTL or BCA.

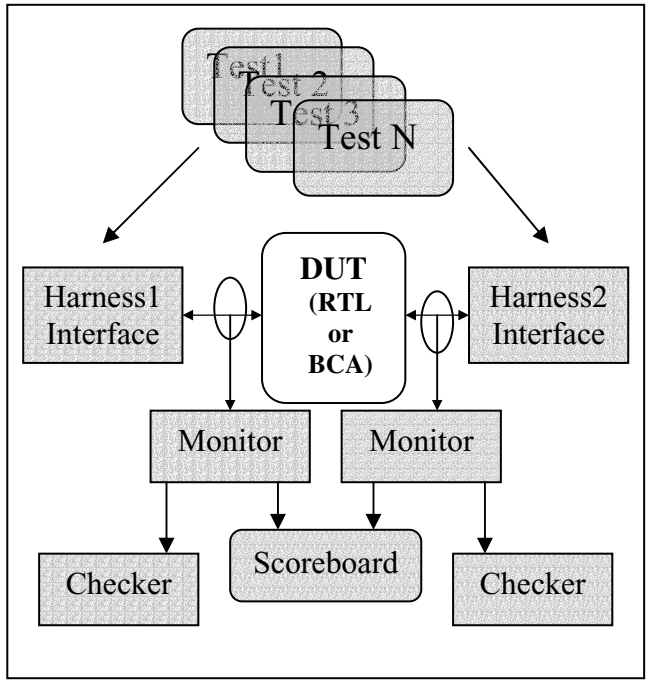

**Figure 2: Generic Testbench Architecture**

The DUT interfaces are connected to eVCs (e Verification Components) written in 'e' code. Each eVC is endowed with BFMs that generate random scenarios, monitors that collect traffic information and checkers that check the correctness of the protocol at the interface. Moreover the scoreboard and specific checkers are required for each DUT to verify the correct behaving of whole according to the verification plan. In order to test particular features of the design, some specific test files need to be developed. Same test file could be run more than one time with a different seed. The aim is to reach a full functional and code coverage rate.

For the STBus interfaces ST has developed an 'e' code generic library called CATG (Checkers and Automatic Test Generation) aimed to test component having STBus interfaces. It contains models of STBus harnesses, monitors, protocol checkers and a scoreboard for data comparison. It has also a detailed functional coverage related to the STBUS. This environment is configurable according to the DUT configuration, in terms of bus size, protocol bus type, pipe size, endianess and some other parameters. It's plugged with DUT, using NCSim's Cadence Simulator. All test bench files except of Specman's 'e' files are compiled and elaborated to give a snapshot to be called by NCSim and Specman tools, at same time.

Specman's environment is plugged to NCSim's VHDL simulator via a provided VHDL wrapper file. This wrapper is called by VHDL test bench file, which contains signals declaration and clocks processes. Except of clock, all signals are driven by eVCs

| SystemC top file node:_top.cpp | VHDL wrapper : node_top.vhd |
|---|---|
| `#include systemc.h`<br>`#include bca_node.h`<br><br>`SC_MODULE(node_top)`<br>`{`<br>`// ports declaration:`<br>`  sc_in<bool> req;`<br>`  sc_in< sc_uint<64> > data;`<br>`   :`<br>`   :`<br><br>`// model declaration`<br>`bca_node *component`<br><br>`// constructor`<br>`SC_CTOR(node_top):`<br>`    req("req"),data("data") {`<br>`    component->req(req);`<br>`    component->data(data);`<br>`    };`<br>`};`<br><br>`// Cadence's specific syntax`<br>`NCSC_MODULE_EXPORT(node_top);` | `LIBRARY ieee;`<br>`USE ieee.std_logic_1164.all;`<br><br>`ENTITY node_top IS`<br>`PORT (`<br>` req : IN STD_LOGIC;`<br>` data : IN STD_LOGIC_VECTOR(63 :`<br>`            downto 0);`<br>`  :`<br>`  :`<br>`);`<br>`END node_top;`<br><br>`-- referring to SystemC model`<br>`ARCHITECTURE SystemC OF node_top IS`<br>`    SystemC : ARCHITECTURE IS`<br>`    "SystemC";`<br>`BEGIN`<br>`END;` |

**Figure 3: Wrappers code for SystemC models**

For what concern the BCA DUT verification an extra work is required in order to connect the model to the common test bench. In fact, CATG library was developed with an old approach not taking into account the port approach, recently introduced by Specman to directly plug SystemC simulator with verification environment. Since many simulators are now able to support SystemC and VHDL design, CATG has been interfaced with SystemC model through VHDL test bench file. However, since VHDL simulator is used, the advantage of having fast SystemC simulator is lost. To interface SystemC simulator and VHDL one, a VHDL wrapper is required, according to Cadence approach. It is similar to the SystemC top file for what concerns signals declaration and module name, and it refers to SystemC model in its architecture (Figure 3). The VHDL test bench is the same as developed for RTL model.

It has the same signals declaration and instantiates the VHDL's wrapper component and the same Specman's wrapper of the RTL model test bench. The compilation process follows Cadence methodology [4].

As verification environment, test benches, wrappers and models's configuration files need to be configured according to model configuration. The regression tool, which is developed internally to run regression flow, generates and compiles these files. It consists on a graphical user interface able to receive configuration parameters. It runs regression tests in batch mode, through generic scripts that are design independent. For each test file associated with the test seed, a verification report and a functional coverage one are generated. Moreover, an associated VCD file, a standard format for waveform recording, is generated so that it can be used later for bus accurate comparison.

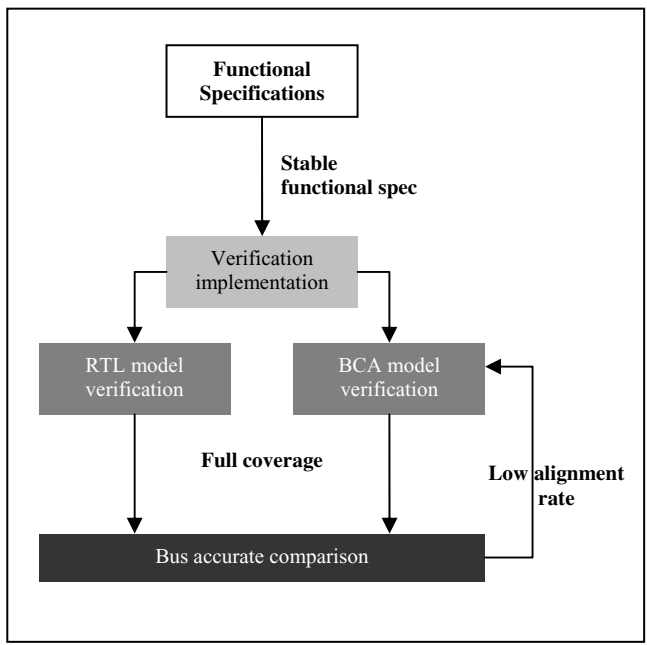

**Figure 4: Common Verification Flow**

The quality of the verification is measured using coverage metrics. Both functional and code coverage must be checked in order to be sure that the design is correctly verified. The functional coverage is built in the common verification environment and it can be obtained in both RTL and BCA models (of course they must be equal running the same tests). The code coverage reflects how the code is exercised and can be applied only in the RTL verification since no tool is able to generate this metrics for SystemC. The code coverage metrics we use are line, branch and statement coverage. Our goal for the verification of the blocks is 100% of the functional coverage defined and 100% of justified code for the line coverage, while in general we accept less for the others code coverage metrics.

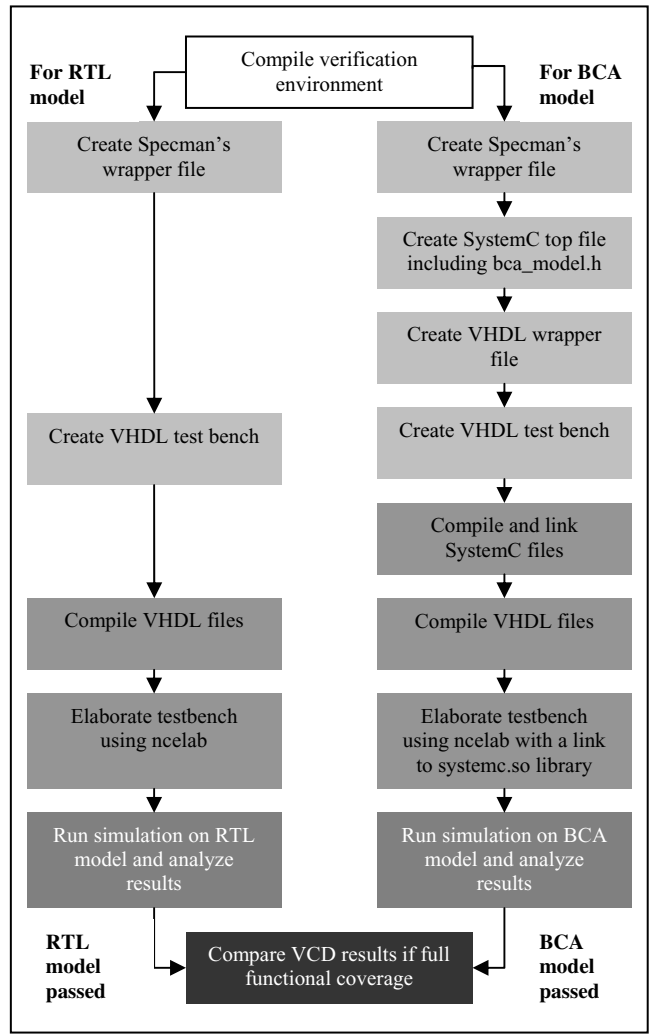

**Figure 5: Common Verification Step by Step**

The quality of the verification is measured using coverage metrics. Both functional and code coverage must be checked in order to be sure that the design is correctly verified. The functional coverage is built in the common verification environment and it can be obtained in both RTL and BCA models (of course they must be equal running the same tests). The code coverage reflects how the code is exercised and can be applied only in the RTL verification since no tool is able to generate this metrics for SystemC. The code coverage metrics we use are line, branch and statement coverage. Our goal for the verification of the blocks is 100% of the functional coverage defined and 100% of justified code for the line coverage, while in general we accept less for the others code coverage metrics.

Having BCA model fully verified with full functional coverage does not guarantee that it's exactly behaving as RTL one, at bus cycle accuracy. This specially happens when the specifications do not constraint signals behavior, so that checkers cannot verify such constraints. A second quality metrics consists on getting bus accurate comparison between both models. STBus Analyzer (STBA), an STBus internal tool, compares signals information at each port level. It is automatically called by the regression tool and it extracts from VCD files, got after regression tests, STBus transaction information. The rate that is calculated at each port level is the number of cycles RTL and BCA signals port are aligned over total number of clock cycles. The targeted value, in order to consider BCA model signed off is 99%.

The Figure 4 shows the complete flow beginning from functional specification down to the bus accurate comparison, while the Figure 5 summarizes steps necessary to implement and to run test bench for both RTL and BCA activities.

## 5. Test case

An example of how this flow has been applied in our team is the verification of the STBus node. The STBus node is the key IP of an STBus interconnect system. It is in fact responsible for performing the arbitration among the requests issued by the initiators of the system, and among the response-requests issued by the targets of the system, and for the routing of the information from the initiators interfaces to the targets interfaces, and vice versa from the targets interfaces to the initiators interfaces.

Supporting either Type 2 or Type 3 STBus protocol, the Node can manage up to 32 initiators and 32 targets and its data interface width varies from 8 to 256 bits. It can have three different architectures: shared bus, full crossbar or partial crossbar. The Node supports 6 arbitration types as Less Recently Used or Latency based. It has an optional programmable port allowing changing the arbitration priority of initiators or targets.

The verification of the Node takes advantage from CATG library. Specific checks, not covered by CATG, have also been developed. The Figure 6 shows a Node with three initiators and two targets. Each harness has its own monitor collecting signals information. STBus protocol interface rules are checked for each port through protocol checkers based on correspondent monitor. In order to test verify data flow integrity between initiators and targets, the scoreboard compares results got form monitors. Harnesses, Monitors, Protocol checkers and scoreboard are all provided by CATG library.

Twelve test cases have been developed to cover the tests of all main features of the node such as out of order traffic or latency based arbitration. They allow initiators to generate semi-random traffic. To force out of order traffic for example, short transactions are sent by one initiator to different targets, having different speed.

The test cases are generic and depend on some HDL parameters. They can be reused for all configurations of the Node.

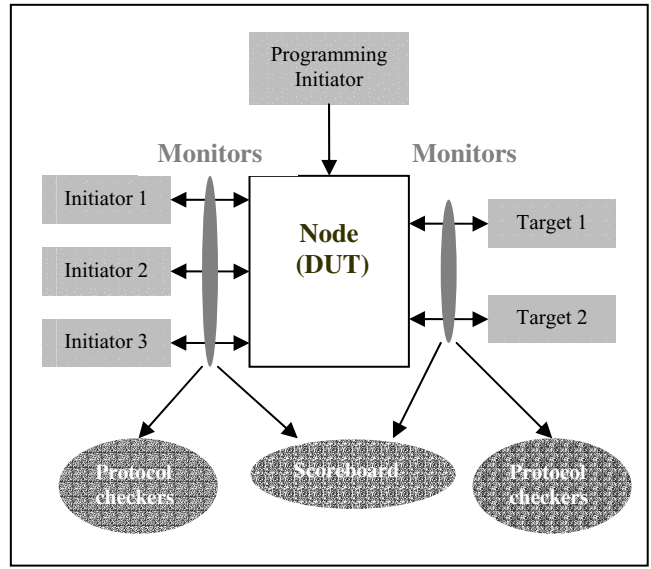

**Figure 6: Node Test bench**

The Regression tool generates VHDL and SystemC files according to used configuration of the Node. HDL Parameters are submitted through a graphical user interface. The tool also launches parallel regression tests on BCA and RTL models. It applies same test cases on both with same seeds. So that it can later, proceed to alignment comparison activity, if all checkers passed.

Since Node has many configurations, regression tool can load text files defining HDL parameters of each of them. It's sufficient to indicate the directory to which the tool has to point.

More than 36 configurations of the Node have been tested. The verification environment permitted to find five bugs on BCA models, not found using old environment of the past flow. It added more confidence on the BCA model to be delivered to STBus customers.

## 6. Conclusion

The verification of the SystemC models is now a key factor to have exact simulation results. A lot of effort has been done in the past in order to improve the RTL verification using specialized verification languages. The goal is to exploit this effort also for the models verification and spend some effort in order to generate a common verification environment reusable for the two views of the design. Moreover it's important to give the visibility to the customers that the adopted strategy in the verification of both models is the same. This has been reached quite easily obtaining good results in terms of time spent in the development of the common environment and in the ability to find bug in both BCA and RTL. In the future this approach will become more and more important so this methodology will be applied in all the STBus activities.

The future availability of the next version of CATG supporting ports approach will make possible a direct interfacing of SystemC simulator with Specman's environment. This should enhance simulation performance. Future including of SystemC Verification in verification flow will be a great opportunity to add TLM (Transaction Level Modeling) development and verification phase in the flow.